\documentclass[conference]{IEEEtran}
\usepackage{cite}     
\usepackage{amsmath,amssymb,amsfonts}
\usepackage{algorithm,algcompatible}
\usepackage{multicol}
\usepackage{soul}
\usepackage{subfloat}
\usepackage{float}
\usepackage{makecell}
\usepackage{multirow}
\usepackage{blkarray}
\usepackage{amsfonts,amsthm,bm}
\usepackage{textcomp}
\usepackage{caption}
\usepackage{subcaption}
\usepackage{stmaryrd}
\usepackage{tabularx}
\usepackage{systeme}
\usepackage{multirow}
\usepackage[table,xcdraw]{xcolor}
\usepackage{tikz}
\usepackage{tikz,pgfplots}
\usepackage{authblk}

\def\BibTeX{{\rm B\kern-.05em{\sc i\kern-.025em b}\kern-.08em
		T\kern-.1667em\lower.7ex\hbox{E}\kern-.125emX}}

\pgfplotsset{ every non boxed x axis/.append style={x axis line style=-},
     every non boxed y axis/.append style={y axis line style=-}}

\begin{document}

\title{Privacy-Preserving Multi-Operator Contact Tracing for Early Detection of Covid19 Contagions}  



\author[1]{\small Davide Andreoletti}
\author[2]{\small Omran Ayoub}
\author[1]{\small Silvia Giordano}
\author[2]{\small Massimo Tornatore}
\author[2]{\small Giacomo Verticale}

\affil[1]{\footnotesize Networking Laboratory, University of Applied Sciences of Southern Switzerland, Manno, Switzerland, Email: \{name.surname\}@supsi.ch }
\affil[2]{\footnotesize Dipartimento di Elettronica, Informazione e Bioingegneria, Politecnico di Milano, Milano, Italy, Email: \{name.surname\}@polimi.it }

\maketitle

\begin{abstract}


The outbreak of coronavirus disease 2019 (covid-19) is imposing a severe worldwide lock-down. Contact tracing based on smartphones' applications (apps) has emerged as a possible solution to trace contagions and enforce a more sustainable selective quarantine. However, a massive adoption of these apps is required to reach the critical mass needed for effective contact tracing. As an alternative, geo-location technologies in next generation networks (e.g., 5G) can enable Mobile Operators (MOs) to perform passive tracing of users' mobility and contacts with a promised accuracy of down to one meter. To effectively detect contagions, the identities of positive individuals, which are known only by a Governmental Authority (GA), are also required. Note that, besides being extremely sensitive, these data might also be critical from a business perspective. Hence, MOs and the GA need to exchange and process users' geo-locations and infection status data in a privacy-preserving manner. In this work, we propose a privacy-preserving protocol that enables multiple MOs and the GA to share and process users' data to make only the final users discover the number of their contacts with positive individuals. The protocol is based on existing privacy-enhancing strategies that guarantee that users' mobility and infection status are only known to their MOs and to the GA, respectively. From extensive simulations, we observe that the cost to guarantee total privacy (evaluated in terms of data overhead introduced by the protocol) is acceptable, and can also be significantly reduced if we accept a negligible compromise in users' privacy.

\end{abstract}


\IEEEpeerreviewmaketitle

\section{Introduction}\label{sec:intro}

Following recent surge of the coronavirus disease (Covid-19) epidemic, various governmental and organizational bodies are expressing strong interest in employing mobile-communication technologies to early detect contagions. 
With the term \lq early detection\rq, we refer to the identification of positive individuals before they show any symptoms. This generally happens during the incubation period of the virus (around $5$ days for Covid-19 \cite{ferretti2020quantifying}), or even during the entire course of the disease. As asymptomatic people unknowingly diffuse the virus, early detection is fundamental to drastically limit virus spread \cite{ferretti2020quantifying}. 

Several smartphone-based apps for early detection are already available \cite{raskar2020apps,cho2020contact,santoro2020covid}. These apps allow a user to know whether she encountered positive individuals or not (e.g., by correlating her mobility with that of known positive cases). In most countries, to comply with strict local privacy regulations, these apps are developed with privacy as a primary design constraint. However, app-based approaches suffer from several drawbacks. First, it is hard to reach the critical mass needed for an effective contact tracing (a typical safe number is $60\%$ of population, a very challenging target \cite{cho2020contact}). In addition, these apps require the continuous use of data acquisition technologies (e.g., the GPS and the Bluetooth) that extensively consume devices' batteries. We also note that people is less likely to keep such apps installed on their smartphones at the very beginning of the epidemics, when early detection is decisive to contain the diffusion of the virus. 

Contact tracing exploiting users' mobility data collected by mobile telecom operators (MOs) is regarded as a promising alternative to app-based solutions \cite{oliver2020mobile}. In the upcoming years, once 5G will have consolidated its penetration, MOs will possess technologies to perform a continuous and accurate tracking of users’ devices. For instance, Ref. \cite{koivisto2017joint} shows that the average accuracy of device positioning in ultra-dense 5G networks will be on the sub-meter order. A great advantage of a passive and continuous positioning is the very limited involvement of the final users. Users are not required to install any application on their smartphones, but only to give an explicit consent to track their position (explicit consent that is currently commonly granted to several apps, such as \cite{raskar2020apps,santoro2020covid}), therefore more easily reaching the critical mass. Even though users (and governments) are becoming more concerned regarding possible violation of the privacy of positioning data by the MOs \cite{scantamburlo2020covid} (e.g., a MO might sell them to third parties), we remark that MOs already estimate subscribers' positions to improve their services \cite{andreoletti2018discovering}, and that the proposed approach guarantees that MOs do not get any sensitive data beyond it.



To effectively obtain an early detection of infections through mobility tracing, in addition to users' mobility data discussed above, the identities of positive individuals, which are only known to a Governmental Authority (GA) through the nation medical institutions, are required, and this is (a possibly even more)-sensitive information that must not be exposed. Therefore, MOs and the GA are required to collect, exchange and process this mobility data (from MO) and infection data (from the GA) in a secure and a privacy-preserving manner. 

In this work, we propose a privacy-preserving protocol that enables GA and MOs to securely share and process users' data, such that each user is guaranteed to be the only person who knows the number of contacts with positive individuals she had (henceforth referred to as user's \textit{score}). The protocol is built on consolidated privacy-enhancing strategies (e.g., secure secret sharing and homomorphic encryption) that guarantee total privacy to users, i.e., the mobility and the infection status of a user are only known to her MO and to the GA, respectively.

This privacy is achieved at an acceptable cost in terms of data overhead exchanged among MOs and GA, as shown from extensive simulations. With slight modifications, the proposed protocol can also be employed i) to make this user discover how many positive people were in her same locations, but not necessarily in her close proximity (e.g., in a pub) and ii) to make the GA know only the identities of the users with a score above a given threshold. The identification of these users would make it possible to more easily stop the diffusion of the virus, but it poses a privacy dilemma and does not comply to several privacy regulations. In this work, we only provide the technical means to realize such identification in a privacy-preserving manner. Note also that, in case the number of these users is high, such procedure requires the exchange of a significant data overhead. However, we also show that this overhead can be heavily reduced at a negligible reduction of users' privacy.

The rest of the paper is structured as follows: in Section \ref{sec:RW} we briefly review some existing approaches for privacy-preserving contact tracing. Section \ref{sec:problem} describes the involved entities and their privacy requirements. We present the building blocks of the proposed protocol and the protocol itself in Sections \ref{sec:protocol_blocks} and \ref{sec:protocol}, respectively. In Section \ref{sec:results} we show some illustrative results obtained by simulation. Finally, Section \ref{sec:conclusions} concludes the paper.

\section{Related Work}\label{sec:RW}

Existing solutions for contact tracing are generally based on smartphone apps of two main types: i) \textit{location-based} (e.g., PrivateKit from MIT \cite{raskar2020apps}) in which user's locations are acquired (e.g., with the GPS technology) and correlated with the locations of positive individuals; ii) \textit{token-based} (e.g., TraceTogether \cite{cho2020contact} and Immuni \cite{santoro2020covid}) that exchange anonymized tokens with smartphones in the proximity of the user (i.e., by exploiting the Bluetooth), and successively match the received tokens with those of known positive persons. 
A user who is tested positive can deliberately share her data (either location or received tokens) with a trusted authority, who then broadcasts it to all the others. Based on this, the app returns if a user has been in contact with a positive individual \cite{raskar2020apps}. As operations are done on users' devices, privacy is mostly preserved, i.e., users' location and contacts are not exposed to the authority. 

However, several privacy issues are still pending. For instance, in location-based apps each user receives the location data of a positive person, whose identity might be obtained from re-identification attacks \cite{cho2020contact,maouche2017ap}. In TraceTogether \cite{cho2020contact} users send their phone numbers and all the received tokens to the authority, which in turn sends a message to those users who met some positive person. As users' contacts are exposed to the authority, this solution would hardly be adopted in countries with strict privacy laws, and several solutions are proposed to solve this issue \cite{cho2020contact}. For example, users might send to the authority only their tokens, and then perform anonymized queries to know if they met some positive person. However, other malicious behaviors are possible given that users obtain the tokens of persons in their proximity. Specifically, a user can craft a query to discover if the person that she met at a given time is positive \cite{cho2020contact}. 

In this work, we exploit consolidated privacy-preserving techniques (e.g., as those employed in \cite{andreoletti2019privacy,andreoletti2019open}) to compute the number of contacts that a user had with positive people, while guaranteeing that users' contacts, locations and infection status are not disclosed to illegitimate parties (see subsection \ref{sec:privacy_requirements} for further details). Differently from token-based solutions that only detect users' proximity, our protocol allows to compute also the number of positive persons within a given place. Unlike existing location-based apps, however, we assume that users' locations are estimated by MOs without any involvement of the users (e.g., using techniques for accurate geo-localization from cellular signals, such as those proposed in \cite{koivisto2017joint}). In this respect, authors in \cite{oliver2020mobile,sathyaprivacy} argue that MOs might play a decisive role in fighting the spreading of a virus, provided that users' privacy is guaranteed. 

\section{Modeling of Involved Entities}\label{sec:problem}


In this Section, we formally define the concept of users' scores, and we describe the role and privacy requirements of the entities involved in their computation. Before doing that, we introduce the concepts of \textit{contact} and \textit{infection status}. We say that two persons $user_{i}$ and $user_{j}$ have a contact iff the distance between them is below a given threshold $th$. We encode this information in the binary variable $c_{ij}^{(t)} = 1$ if $Dist(loc_{i}^{(t)},loc_{j}^{(t)}) < th$, and $0$ otherwise, where $loc_{i}^{(t)}$ and $loc_{j}^{(t)}$ refer to the geo-location (e.g., latitude and longitude) of $user_{i}$ and $user_{j}$ at time $t$, respectively, while $Dist$ is a measure of geographical distance. Concerning the infection status, we then introduce the binary variable $s_{i}^{(t)} = 1$ in case $user_{i}$ is considered positive at time $t$, and $0$ otherwise. $Score_{i}$ is the number of contacts that $user_{i}$ has, during a given period of time, with positive individuals. In formulas:

\begin{equation}\label{eq:score}
Score_{i} = \sum_{t}\sum_{j:c_{ij}=1} s^{(t)}_{j}
\end{equation}

Similarly, $Score^{(Loc)}_{i}$ is the number of positive persons that were in a certain location $Loc$ at the same time of $user_{i}$, and is computed as $Score^{(Loc)}_{i} =\sum_{\substack{j \in Loc}} s^{(t)}_{j}$, where the considered locations are assumed to be chosen by $user_{i}$ herself.

\subsection{Role of Involved Entities}\label{sec:entities}

The GA is an entity established by the government to monitor the infection status of individuals within a certain region and, specifically, to collect from medical institutions the identities of positive individuals willing to share this data.

MOs are instead telecom companies that provide mobile connectivity within the considered region. Without loss of generality, we assume that each user is served by only one MO, and that the whole area is covered by all the MOs. Then, we also assume that MOs estimate the locations of their users at time $t$, i.e., $\hat{loc}_{i}^{(t)}, \forall i$ from cellular signals received by users' devices (e.g., as done in \cite{koivisto2017joint}).

\subsection{Privacy Requirements and Security Models}\label{sec:privacy_requirements}

We assume that the GA and the MOs are \textit{honest-but-curious}, i.e., they honestly execute the protocol but also try to violate other parties' privacy from the received data. Privacy requirements for each type of data are illustrated below.

\subsubsection{Users' Locations} estimates of a user's locations should only be known to her MO. 

\subsubsection{Users' Contacts}

information regarding contacts between two users should only be known to their MOs. In addition, if these users are subscribers of different MOs, each MO should not know anything neither about the identity of the other MO's user, nor about the number of contacts between its users and any other user of its competitors (e.g., how many contacts $user_{i}$ and $user_{j}$ have during a given period).    
\subsubsection{Users' Infection Status and Scores}

The infection status of a user should only be known to the GA and to the user herself (say $user_{i}$). $Score_{i}$ and $Score_{i}^{(Loc)}$ should only be known to $user_{i}$, except when $Score_{i}$ is greater than a threshold $\chi$. In this case, $Score_{i}$ and the identity of $user_{i}$ might also be known to the GA (see subsection \ref{sec:communication} for the details).

\section{Building Blocks of the Privacy-Preserving Protocol}\label{sec:protocol_blocks}

\subsection{Existing Privacy-Preserving Building Blocks}\label{sec:background_privacy}

\subsubsection{Shamir Secret Sharing}

A Shamir Secret Sharing (SSS) scheme \cite{shamir1979share} allows to securely distribute a secret $s$ among a set of participants in such a way that $s$ can only be recovered if a sufficient number of them cooperate. The piece of secret $s$ that each participant receives is called \textit{share}, and it is referred to as $\llbracket s \rrbracket$. In this work, we employ a $(2,2)$ SSS, i.e., $s$ is reconstructed only if $2$ out of the $2$ considered participants cooperate. 
SSS has several homomorphic properties, i.e., each participant can perform several operations on the shares that result in the same operations over the corresponding secrets (e.g., linear combinations). Then, participants can compute $\llbracket s_{1}\cdot s_{2} \rrbracket$ using the \textbf{Mult} protocol presented in \cite{turban2014secure}, or they can use the \textbf{EQ} and \textbf{Comp} protocols \cite{turban2014secure} to perform the equality check and the comparison operations. In the latter, participants input their shares $\llbracket s_{1} \rrbracket$ and $\llbracket s_{2} \rrbracket$ and obtain the share $\llbracket b_{eq} \rrbracket$ (resp., $\llbracket b_{ge} \rrbracket$), where $b_{eq} = 1$ (resp., $b_{ge} = 1$) iff $s_{1} = s_{2}$ (resp., $s_{1} \geq s_{2}$) and $0$ otherwise. 



\subsubsection{Paillier Cryptosystem}\label{sec:paillier}

Paillier \cite{paillier1999public} is a secure cryptosystem with the following properties: i) it is asymmetric, i.e., anyone can encrypt a message, but only the owner of the private key can decrypt it; ii) it is probabilistic, i.e., two encryptions of the same plaintext yield different ciphertexts and iii) it is homomorphic with respect to the summation of two ciphertexts (computed as $Enc(m_{1} + m_{2}) = Enc(m_{1})\cdot Enc(m_{2})$) and to the product between a ciphertext and a plaintext (computed as $Enc(m_{1}\cdot m_{2}) = Enc(m_{1})^{m_{2}}$).

\subsection{New Privacy-Preserving Primitives based on SSS}

\subsubsection{\textbf{Secure Square Distance}}\label{sec:secsqudist}
the \textit{Secure Square Distance} module takes in input the shares of the coordinates of points $i$ and $j$, i.e., $\llbracket x_{i} \rrbracket, \llbracket y_{i} \rrbracket, \llbracket x_{j} \rrbracket, \llbracket y_{j} \rrbracket$ and returns $\llbracket d_{ij}^{2} \rrbracket$, where $d_{ij}$ is the euclidean distance between these points. 
This module is based on the Mult subroutine. 

\subsubsection{\textbf{ObliviousTransfer}}\label{sec:SSS-baed_OT}

the ObliviousTransfer module (OT) allows a sender to deliver some data to a receiver without knowing which data has been transmitted. OT inputs i) a set of $2N$ shared elements arranged into a table with $N$ rows and two columns (namely, \textit{attribute} and \textit{value}) and ii) the share $\llbracket attribute_{x} \rrbracket$. This module is based on the Mult and EQ subroutines and outputs the share $\llbracket value_{i} \rrbracket$ if the attribute at row $i$ is equal to $attribute_{x}$, and $\llbracket 0 \rrbracket$ otherwise. This value is computed as $\llbracket value_{i} \rrbracket = \sum_{j=1}^{N} \llbracket eq_{jx} \cdot value_{j} \rrbracket $, where $eq_{jx} = 1$ if $attribute_{x} = attribute_{j}$, and $0$ otherwise. 

\section{The privacy-preserving protocol}\label{sec:protocol}

The proposed protocol works in three main phases, namely \textit{contact tracing}, \textit{score computation} and \textit{communication with users}. We describe these phases in the following subsections. We refer to the generic users $user_{i}$ and $user_{j}$ as subscribers of $MO_{k}$ and $MO_{k\prime}$, respectively, but the described operations are valid for each user and MO.

\subsection{Privacy-Preserving Contact Tracing}\label{sec:contact_tracing}

In this phase, $MO_{k}$ obtains the binary value $c_{ij}$ encoding the information about its generic $user_{i}$'s contacts, $\forall i$. 
Firstly, $MO_{k}$ estimates the current location of $user_{i}$, i.e., $(\hat{lat}_{i}^{(t)},\hat{long}_{i}^{(t)})$ by analyzing cellular signals coming from her device \cite{koivisto2017joint}. 
From this data, the $MO_{k}$ can independently assess the contacts among its subscribers, but not with other MOs' users (since a free exchange of users' mobility data is prohibited by the considered privacy requirements). Hence, we propose to perform the privacy-preserving computation of $c_{ij}$ as follows.

\begin{figure}[h!]{ 
\centering 
{
\includegraphics[width=8cm,height=4cm]{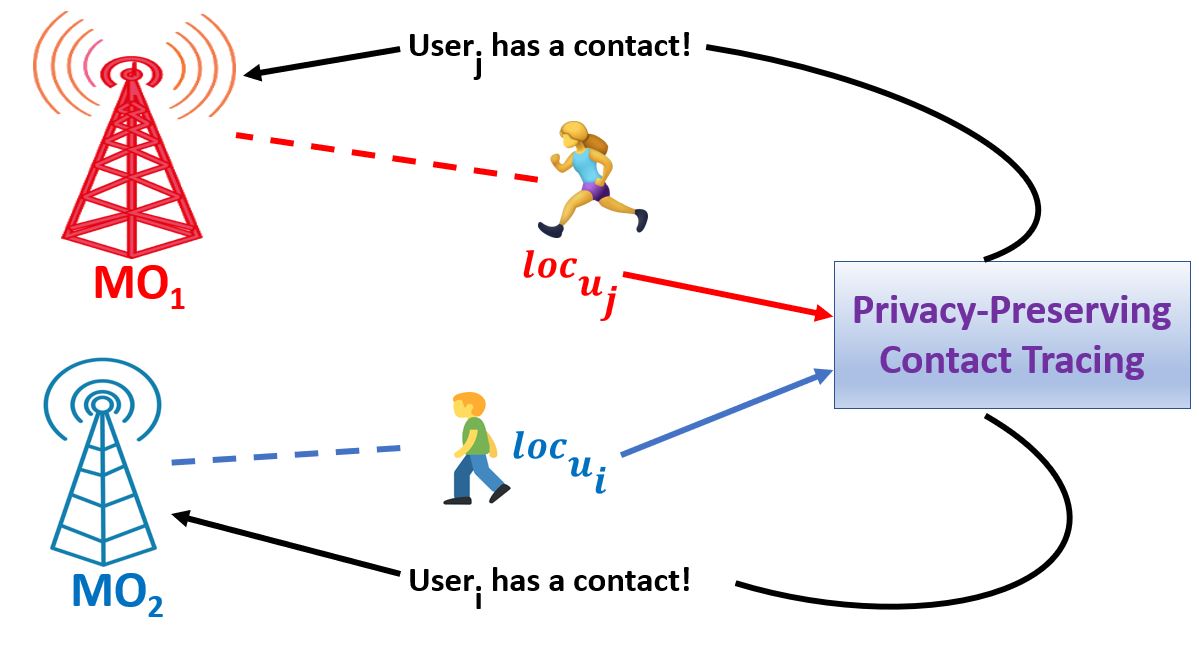}
}
\caption{Positioning process and privacy-preserving contact tracing performed by a pair of MOs
}\label{fig:contact_tracing}
}
\end{figure}

$MO_{k}$ and $MO_{k\prime}$ compute the projections of their users' estimated positions on an euclidean plane (e.g., $\hat{x}^{(t)}_{i}, \hat{y}^{(t)}_{i}$) and exchange these values among them in form of secret shares. Then, they execute the Secure Square Distance module and obtain $\llbracket d^{2}_{ij}\rrbracket$, being $ d^{2}_{ij}$ the squared euclidean distance between the generic $user_{i}$ and $user_{j}$. The Comp module is then employed to compare $\llbracket d^{2}_{ij} \rrbracket$ with the threshold $\llbracket th^{2} \rrbracket$ and obtain $\llbracket c_{ij}\rrbracket$. The MOs finally exchange these shares and recover the secret $c_{ij}$ (that is $1$ if $user_{i}$ and $user_{j}$ has a contact, and $0$ otherwise). A representation of this phase is depicted in Fig. \ref{fig:contact_tracing}. 

\subsection{Secure Computation of Users' scores}\label{sec:score_computation}

In this phase, $MO_{k}$ securely computes the score values of $user_{i}$. To do so, the GA sends to $MO_{k}$ the infection status (in encrypted form) of $user_{i}$ during a considered period (e.g., in the last day), i.e., $Enc_{GA}(s_{i}^{(t)}), \forall t$. At each time instant of the considered period, $MO_{k}$ and $MO_{k\prime}$ obtain $c_{ij}^{(t)}$ as described in the previous subsection. If this value is $1$ (i.e., there is a contact between these users at time $t$), $MO_{k}$ and $MO_{k\prime}$ exchange with each other $Enc_{GA}(s^{(t)}_{i})$ and $Enc_{GA}(s^{(t)}_{j})$. Then, $MO_{k}$ computes $Enc_{GA}(Score_{i})$ by homomorphically executing the summation in Eq. \ref{eq:score}. Similarly, $MO_{k}$ computes $Score_{i}^{(Loc)}$ by homomorphically summing the encrypted infection status of all users within area $Loc$ at a given time, which are asked to all the remaining MOs. 

The obtained data are then arranged by $MO_{k}$ in a table that we represent in Table \ref{tab:mok_data}. Such table has $N_{k}$ rows (one of each subscriber of $MO_{k}$) and three columns, which are \textit{Index}, \textit{Identity} and \textit{Score}. The first refers to the index of the row at which a certain user's data is stored. Without loss of generality, we assume that $user_{i}$'s data is stored at the $i$-th row. The second one stores the identities of the users (e.g., anything allowing to univocally identify them, such as full names and telephone numbers). The third represents the $Score$ values of users in encrypted form.

\begin{table}[]
\centering 
\caption{Data of subscribers of $MO_{k}$}
\label{tab:mok_data}
\begin{tabular}{|c|c|c|}
\hline
\textbf{Index} & \textbf{Identity}                   & \textbf{Score} \\ \hline
1                               & $Id_{1} = Name_{1}||PhoneNumber_{1}$            & $Enc_{GA}(score_{1})$           \\ \hline
$...$                           & $...$                                                  & $...$                           \\ \hline
$i$                             & $Id_{1} = Name_{i}||PhoneNumber_{i}$            & $Enc_{GA}(score_{i})$           \\ \hline
$...$                           & $...$                                                  & $...$                           \\ \hline
$N_{k}$                         & $Id_{N_{k}} = Name_{N_{k}}||PhoneNumber_{N_{k}}$ & $Enc_{GA}(score_{N_{k}})$       \\ \hline
\end{tabular}
\end{table}

\subsection{Communication with users}\label{sec:communication}

In this phase, we show how to distribute users' scores only to the legitimate entity (i.e., either the user herself or the GA). We consider the scenarios of  \textit{User-Triggered Communication} and \textit{GA-Triggered Communication}. In the former, scores are requested by $user_{i}$ herself, and are kept secret to any other entity. In the latter, the GA identifies only the users with a score greater than a given threshold $\chi$. 


\subsubsection{User-Triggered Communication}\label{sec:citizen_triggered}

\begin{figure}[t!]{ 
\centering 
{
\includegraphics[width=7cm,height=3.5cm]{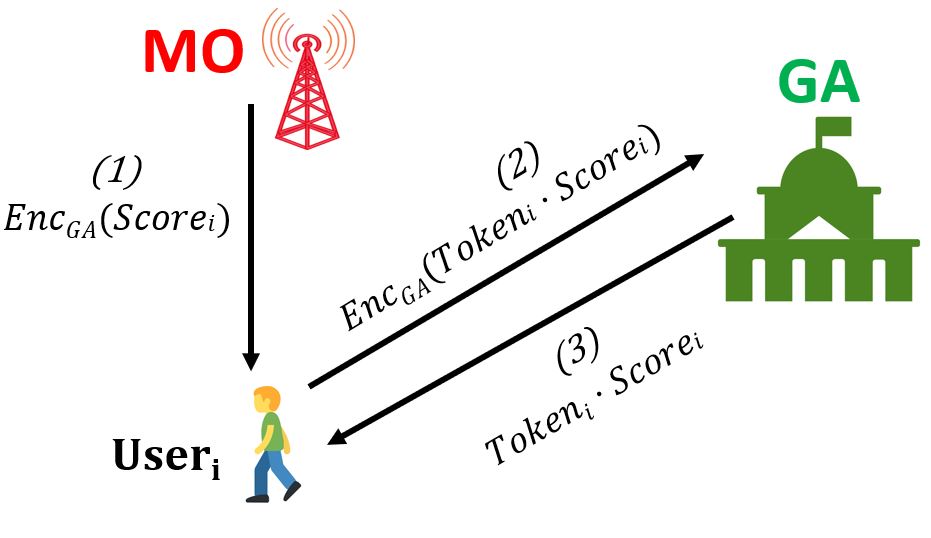}
}
\caption{Representation of the User-triggered communication
}\label{fig:user_triggered}
}
\end{figure}

$user_{i}$ directly asks to $MO_{k}$ the values $Enc_{GA}(score_{i})$ and $Enc_{GA}(score^{(Loc)}_{i})$, for any location she is interested in. Then, $user_{i}$ exploits the homomorphic properties of the Paillier cryptosystem to compute $Enc_{GA}(Score_{i}\cdot Token_{i})$, where $Token_{i}$ is a random value known only to her. $Enc_{GA}(Score_{i}\cdot {Token_{i}})$ is then sent to the GA, which deciphers it and sends $Score_{i}\cdot {Token_{i}}$ back to $user_{i}$. Finally, $user_{i}$ removes the mask $Token_{i}$ and obtains $Score_{i}$. A similar computation is performed to obtain $score^{(Loc)}_{i}$. We represent this phase in Fig. \ref{fig:user_triggered}.

\subsubsection{GA-Triggered Communication}\label{sec:ga-triggered}

$MO_{k}$ sends to the GA $index_{x}$ and $Enc_{GA}(score_{x}), \forall x$. Then, the GA deciphers $Enc_{GA}(score_{x})$ and obtains $(index_{x},score_{x}), \forall x$. In case $\exists i : score_{i} \geq \chi$, the GA and $MO_{k}$ jointly execute the OT subroutine described in subsection \ref{sec:SSS-baed_OT}. 
To do so, $MO_{k}$ sends to the GA $\llbracket index_{x} \rrbracket$ and $\llbracket identity_{x} \rrbracket, \forall x$, while the GA sends to $MO_{k}$ $\llbracket index_{i} \rrbracket$. With these values in input, the OT module returns to $MO_{k}$ and to the GA their shares $\llbracket identity_{i} \rrbracket$. 
Finally, $MO_{k}$ sends its share $\llbracket identity_{i} \rrbracket_{MO_{k}}$ to the GA, which combines it with $\llbracket identity_{i} \rrbracket_{GA}$ and recover the identity of $user_{i}$. In the next subsection, we show how the proposed protocol fulfills the considered privacy requirements under the honest-but-curious security model. 

\subsection{Fulfillment of Privacy Requirements}


\subsubsection{Users' Locations}during the contact tracing phase, estimated users' locations are distributed among pairs of MOs as secret shares. As SSS is proven information-theoretic secure \cite{shamir1979share}, no information about locations is obtained from the single shares owned by each MO. 

\subsubsection{Users' Contacts}at each execution of the contact tracing phase, pairs of MOs distribute to each other new shares of their users' locations. This prevents a leakage of users' identities (which cannot be inferred from locations' shares), as well as from counting the number of contacts between two users. Then, during the score computation phase, $MO_{k\prime}$ can homomorphically compute $Enc_{GA}(s_{j} + 0)$ (which yields a different ciphertext without altering the hidden infection status), in such a way that $MO_{k}$ cannot count the number of contacts between $user_{i}$ and $user_{j}$.


\subsubsection{Users' Infection Status and Scores} $MO_{k}$ computes $user_{i}$'s scores by performing homomorphic summations on values encrypted by the GA but, since it does not know the private encryption key, it does not discover any plaintext. Then, in the user-triggered communication scenario, $user_{i}$ sends $Enc_{GA}(s_{i}\cdot Token_{i})$ to the GA. As the latter does not know $Token_{i}$, it cannot obtain the actual values of the scores. Finally, in the GA-triggered communication scenario, GA and $MO_{k}$ execute the OT module. From this execution, the GA learns the identity of $user_{i}$ and $MO_{k}$ learns nothing. As the GA is considered a honest-but-curious entity, we assume that it executes the OT module only to identify users with the highest chance to be positive (i.e., if $score_{i} \geq \chi$). Clearly, the GA might execute this module regardless of the value of $score_{i}$ and learn the identity and scores of all the users. In the next subsection, we discuss a possible extension of the protocol to cope with this malicious behaviour of the GA.

\subsection{Extension of the protocol for dishonest participants}


We now describe how the protocol can be improved to address two malicious schemes. In the first one, the GA tries to obtain the identity of $user_{i}$ when $Score_{i} < \chi$. The proposed solution works as follows: $MO_{k}$ selects two random variables $\tau_{1}$ and $\tau_{2}$ and computes $Enc_{GA}(\tau_{1}\cdot score_{x} + \tau_{2}), \forall x$. These values are then sent to the GA in form of secret share, i.e., $\llbracket Enc_{GA}(\tau_{1}\cdot score_{x} + \tau_{2})\rrbracket, \forall x$ and given in input to the OT module. From its execution, $MO_{k}$ and the GA obtain $\llbracket Enc_{GA}(\tau_{1}\cdot score_{i} + \tau_{2}) \rrbracket$. $MO_{k}$ sends its share to the GA, which can then recover the secret $Enc_{GA}(\tau_{1}\cdot score_{i} + \tau_{2})$ and, from it, the plaintext $\tau_{1}\cdot score_{i} + \tau_{2}$. Finally, the GA sends to $MO_{k}$ both $\tau_{1}\cdot score_{i} + \tau_{2}$ and $score_{i}$. Since the GA never obtains the values $\tau_{1}$ and $\tau_{2}$, it cannot counterfeit a $score_{i} \geq \chi$ and a corresponding valid $\tau_{1}\cdot score_{i} + \tau_{2}$. $MO_{k}$ detects a cheat if $score_{i} < \chi$ or the actual $\tau_{1}\cdot score_{x} + \tau_{2}$ cannot be computed from $score_{i}$. If the GA does not cheat, the OT module is executed again as previously described, and the GA obtains $identity_{i}$.

In the second malicious scheme, $MO_{k}$ counterfeits the encryption infection status of $user_{i}$. To address this issue, the GA sends to $MO_{k}$ the infection status of users multiplied by a constant, e.g., $Enc_{GA}(s_{i}\cdot Token_{GA})$, where $Token_{GA}$ is known to the GA only. The GA detects a cheat if the ciphered score computed by $MO_{k}$ does not decrypt to a multiple of $Token_{GA}$ (i.e., $Score_{i}\cdot Token_{GA}$).

\section{Illustrative Numerical Results}\label{sec:results}

\subsection{Simulation Settings}

We perform our experiments considering a population of $N = 1.5$ millions users, whose mobility is traced every $20$ seconds within an overall period of $1$ hour. 
The initial position of the generic $user_{i}$ is given by $x_{i} = R_{i}cos(\theta_{i}),y_{i} = R_{i}sin(\theta_{i})$, being $R_{i}$ and $\theta_{i}$ two random variables that follow the Gaussian distribution (with zero mean and standard deviation equal to $  3800m$) and the uniform distribution defined over $[0,2\pi]$, respectively. Users move following the Gauss-Markov model \cite{camp2002survey} ($40\%$ of them at an average speed of $0.01$m/s, $40\%$ at $1$m/s and the remaining $20\%$ at $14$m/s). The region occupied by the population is $1900km^{2}$ large, and is covered by $K \in [2,5]$ MOs, who have the same number of subscribers $\frac{N}{K}$. $1\%$ of the whole population is assumed to be currently positive. 

\begin{figure}[b!]
\centering
\input{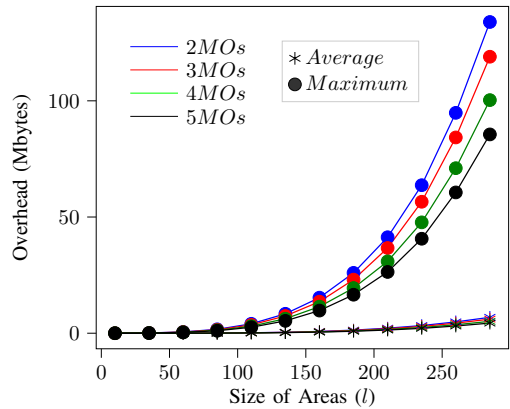}
    \caption{Overhead of data exchanged by each MO, during the contact tracing phase, within areas of size $l$}
    \label{fig:overhead_contact_tracing}
\end{figure}

\begin{figure}[h!]
\centering
\input{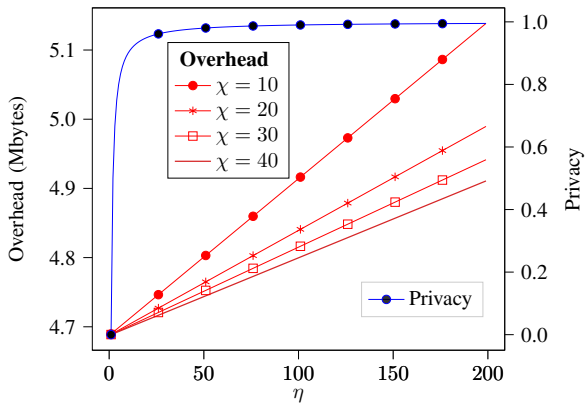}
    \caption{Trade-off between data overhead and privacy with varying the range $\eta$, for several values of the threshold $\chi$ and $K=5$}
    \label{fig:trade_off}
\end{figure}

\subsection{Data Overhead}\label{sec:overhead}

We now show the overhead generated in each phase of execution of the protocol, being $b$ the bit-length of the shares exchanged by participants (in our simulations $b=25$ bits).

\begin{table*}[t!]
\centering
\vspace*{0.325cm}
\caption{Time needed to perform the contact tracing phase within a given subarea}
\begin{tabular}{c|c|c|c|c|c|c|c|c|c|c|c|c|}
\cline{2-13}
\multirow{2}{*}{}                & \multicolumn{12}{c|}{Size of Subareas $l$ (meters)}                                                                                  \\ \cline{2-13} 
                                 & 10     & 35     & 60      & 85      & 110      & 135      & 160      & 185     & 210     & 235     & 260     & 285      \\ \hline
\multicolumn{1}{|c|}{Avg Timing (seconds)} & $2\cdot10^{-5}$ & $2.4\cdot10^{-2}$ & $1.8\cdot10^{-1}$ & $6.7\cdot10^{-1}$ & $1.7$ & $3.8$ & $7.2$ & $12.4$ & $20.2$ & $30.8$ & $45.7$ & $64.1$  \\ \hline
\multicolumn{1}{|c|}{Max Timing (seconds)} & $7\cdot{10^{-2}}$   & $1$ & $5.1$  & $15.6$ & $37.7$  & $78.3$  & $143.5$  & $244.1$ & $388.5$ & $599.1$ & $892.7$ & $1260.6$ \\ \hline
\end{tabular}
\label{sec:time_overhead}
\end{table*}

\subsubsection{Contact tracing phase}



We assume that two users have a contact if their distance is below $th = 2$m. The overhead generated to evaluate if there is a contact is $18b^{2} + 10b$. The number of these evaluations depend on the number of users currently located within a given area, which in turns depends on its size. To avoid comparisons among users with a negligible probability to meet, we assume that contacts are searched within non-overlapping squares of size $l$. In Fig. \ref{fig:overhead_contact_tracing}, we show the average and maximum overhead generated by each MO to execute the contact tracing phase over an area of size $l \in \{10,35,60,...,285\}$ meters. 
From this figure, we observe a super-linear increase of both the average and maximum overhead per area with increasing $l$. We also notice that the overhead is higher when decreasing the number of involved MOs $K$. While the average overhead is always less than $6.8$ Mbytes, the maximum overhead grows significantly with $l$. As an example, when $K = 2$ the maximum overhead goes from $  0.01$ to $134$ Mbytes when $l$ goes from $10$ to $285$ meters. 

\subsubsection{Score Computation phase} With 4096 bit-long ciphertexts \cite{jost2015encryption}, the overhead at each execution of the score computation phase (every $1$ hour in our simulations) is $768$ Mbytes from the GA to the MOs (i.e., obtained by delivering the infection status of users), and $153$ Mbytes among MOs (i.e., obtained by exchanging the infection status of their users in case of contact). 

\subsubsection{Communication with users phase} The overhead generated in the user-triggered communication is negligible (i.e., $  1.5$ Kbytes/user). On the other hand, the GA-triggered communication generates a total overhead of $14N_{\chi}N_{k}b^{2} + 2N_{\chi}N_{k}b + Nb$  bits, where $N_{\chi}$ is the number of users whose score is $\geq \chi$. 
For instance, for $K = 5$ and $\chi = 10$ the overhead is $4650$ Mbytes. Although this value can be considered acceptable, we note that it would be much higher if longer periods and a higher number of users were considered. To reduce this overhead, the GA sends to the MO both the share $\llbracket index_{i} \rrbracket$ and a range $[index_{i}-\eta_{-},index_{i}+\eta_{+}]$ that indicates the rows of Table \ref{tab:mok_data} in which the identity of $user_{i}$ should be searched. Since the GA-triggered communication is issued only for users with a score $\geq \chi$, the MO discovers that one among its users with index $\in [index_{i}-\eta_{-},index_{i}+\eta_{+}]$ in Table \ref{tab:mok_data} has a higher-then-average chance to be positive. Hence, there is a trade-off between overhead and $user_{i}$'s privacy. We measure privacy as the probability that her MO discovers that $user_{i}$ has a score $\geq \chi$, i.e., $privacy_{i} = 1 - \frac{1}{\eta}$, where $\eta = \eta_{-} + \eta_{+}$. In Fig. \ref{fig:trade_off}, we show the trade-off between overhead and privacy with varying $\eta \in [1,200]$, for $\chi \in [10,20,30,40]$. From this figure, we observe that a very high level of privacy can be reached at a remarkable reduction of the overhead. For instance, for $\chi = 10$ the overhead drops from $4650$ to $5$ Mbytes if we accept $99.5\%$ of the total privacy. 



\subsection{Computational Time}

In Table \ref{sec:time_overhead} we show the average and maximum time needed to compute the contacts among users, for several values of $l$. We note that $l$ should not exceed $85$ meters to allow a sampling of users' mobility every $20$ seconds. Then, the GA-triggered communication for a single user takes $\tau \cdot \eta$, with $\tau =6 ms$ on a Intel Core I7 computer. When $K=5$, the identities of users with score $\geq 10$ are obtained in $66$ minutes if total privacy is considered (i.e., if $\eta = 3\cdot 10^{5}$). This value drops to $2.64$ seconds if the $99.5\%$ of privacy is considered sufficient.

\section{Conclusion}\label{sec:conclusions}

We proposed a privacy-preserving protocol that enables a GA (owning users' infection status) and several MOs (owning accurate estimations of users' positions) to compute the number of contacts that users have with positive persons, during a considered period. The protocol guarantees that such measure is only obtained by the legitimate user, and that her infection status and mobility data are known, respectively, only to her MO and to the GA. The protocol can also be employed i) to make a user know the number of positive people who stayed in her same area (even thought not in close contact with her) and ii) to make the GA discover the identities only of users with the highest chance to be positive. We evaluated the cost of privacy in terms of overhead generated by the protocol. From extensive simulations, we observed that the overhead is acceptable, and can further be reduced at a negligible reduction of users' privacy.

\bibliographystyle{IEEEtran}
\bibliography{main}

\end{document}